\documentclass[10pt,twocolumn,letterpaper]{article}

\usepackage{wacv}
\usepackage{times}
\usepackage{graphicx}
\usepackage{enumerate}
\usepackage{amsthm}
\usepackage{url}
\usepackage{balance}

\usepackage{graphicx}
\usepackage{booktabs}
\usepackage{multirow}
\usepackage{xspace}
\usepackage{amsmath,amssymb}
\usepackage{color}
\usepackage[pagebackref=false,breaklinks=true,colorlinks,bookmarks=false,urlcolor=blue,linkcolor=blue,citecolor=blue,pdftitle={Classification of Human Epithelial Type 2 Cell Indirect Immunofluoresence Images via Codebook Based Descriptors},pdfauthor={Arnold Wiliem, Yongkang Wong, Conrad Sanderson, Peter Hobson, Shaokang Chen, Brian C. Lovell}]{hyperref}

\def\Vec#1{{\boldsymbol{#1}}}
\def\Mat#1{{\boldsymbol{#1}}}

\def\argmin#1#2{{\underset{#1}{\operatorname{arg\,min}}~#2}}

\wacvfinalcopy
\def\wacvPaperID{xx}

\title
  {
  Classification of Human Epithelial Type 2 Cell\\
  Indirect Immunofluoresence Images via Codebook Based Descriptors
  }

\author
  {
  {\it Arnold~Wiliem, Yongkang~Wong, Conrad~Sanderson, Peter~Hobson, Shaokang~Chen, Brian~C.~Lovell}\\
  ~\\
  NICTA, PO Box 6020, St Lucia, QLD 4067, Australia \thanks{{\bf Published~in:} IEEE Workshop on Applications of Computer \mbox{Vision}, pp.~95--102, 2013. \href{http://dx.doi.org/10.1109/WACV.2013.6475005}{http://dx.doi.org/10.1109/WACV.2013.6475005}}\\
  The University of Queensland, School of ITEE, QLD 4072, Australia \\  
  Sullivan Nicolaides Pathology, Australia
  }

\begin{document}
\maketitle
\thispagestyle{empty}

\begin{abstract}
\vspace{-2ex}

\noindent
The Anti-Nuclear Antibody (ANA) clinical pathology test is commonly
used to identify the existence of various diseases.
A hallmark method for identifying the presence of ANAs 
is the Indirect Immunofluorescence method on Human Epithelial (\mbox{HEp-2}) cells,
due to its high sensitivity and the large range of antigens that can be detected.
However, the method suffers from numerous shortcomings, such as being subjective as well as time and labour intensive.
Computer Aided Diagnostic (CAD) systems have been developed to address these problems,
which automatically classify a \mbox{HEp-2} cell image into one of its known patterns (eg.,~speckled, homogeneous).
Most of the existing CAD systems use handpicked features to represent a \mbox{HEp-2} cell image,
which may only work in limited scenarios.
In this paper, we propose a cell classification system comprised of a dual-region codebook-based descriptor, 
combined with the Nearest Convex Hull Classifier. 
We evaluate the performance of several variants of the descriptor on two publicly available datasets:
ICPR \mbox{HEp-2} cell classification contest dataset and the new \mbox{SNPHEp-2} dataset.
To our knowledge, this is the first time codebook-based descriptors are applied and studied in this domain. 
Experiments show that the proposed system has consistent high performance
and is more robust than two recent CAD systems. 

\end{abstract}

\vspace{-3ex}
\section{Introduction}
\vspace{-1ex}

In recent years, there has been increasing interest in employing image analysis 
techniques for various routine clinical pathology tests~\cite{Gurcan2009,Hiemann2009,Khutlang2010}. 
Results produced by these techniques can be incorporated into subjective analysis done by scientists,
leading to test results being more reliable and consistent across laboratories~\cite{Hiemann2009}. 

\begin{figure}[!t]
  \begin{minipage}{1.0\columnwidth}
    \centering
    \hfill
    \begin{minipage}{0.45\textwidth}
      \centering
      \includegraphics[width=1.0\textwidth]{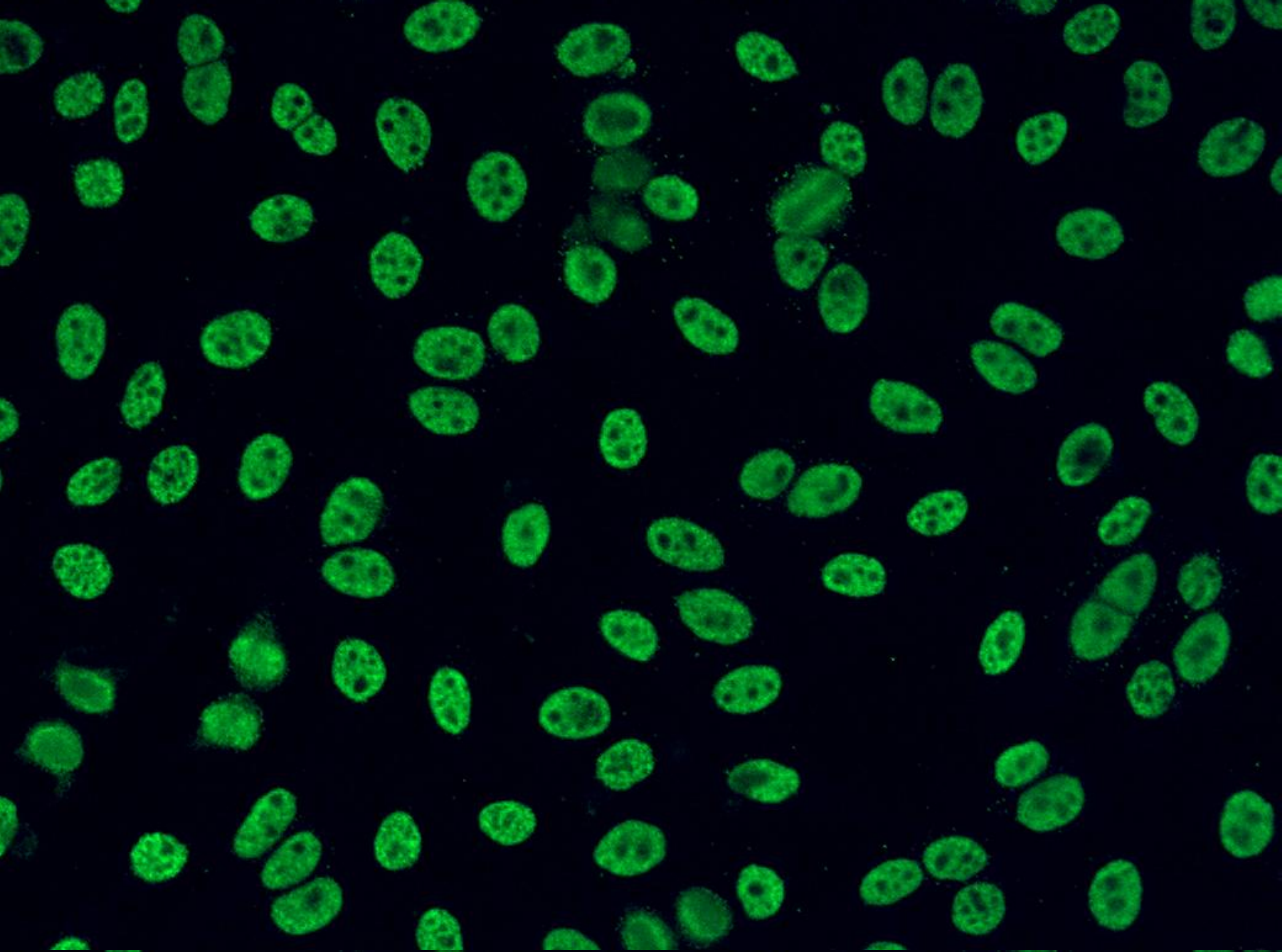}
      {\small speckled}
    \end{minipage}
    \hfill
    \begin{minipage}{0.45\textwidth}
      \centering
      \includegraphics[width=1.0\textwidth]{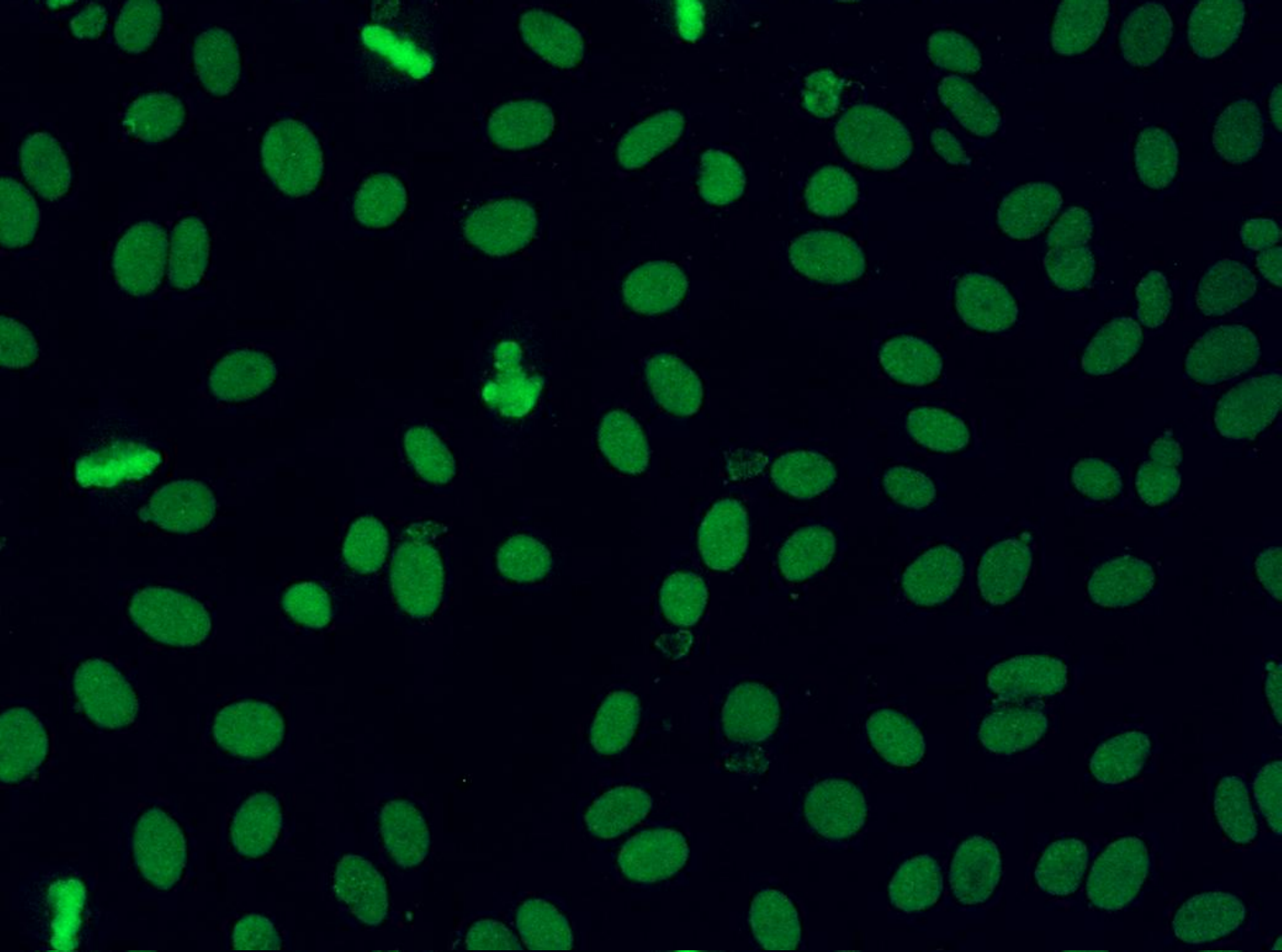}
      {\small homogeneous}
    \end{minipage}    
    \hfill
    ~
  \end{minipage}    
  
  \vspace{1ex}
  
  \begin{minipage}{1.0\columnwidth}
    \centering
    \hfill
    \begin{minipage}{0.45\textwidth}
      \centering
      \includegraphics[width=1.0\textwidth]{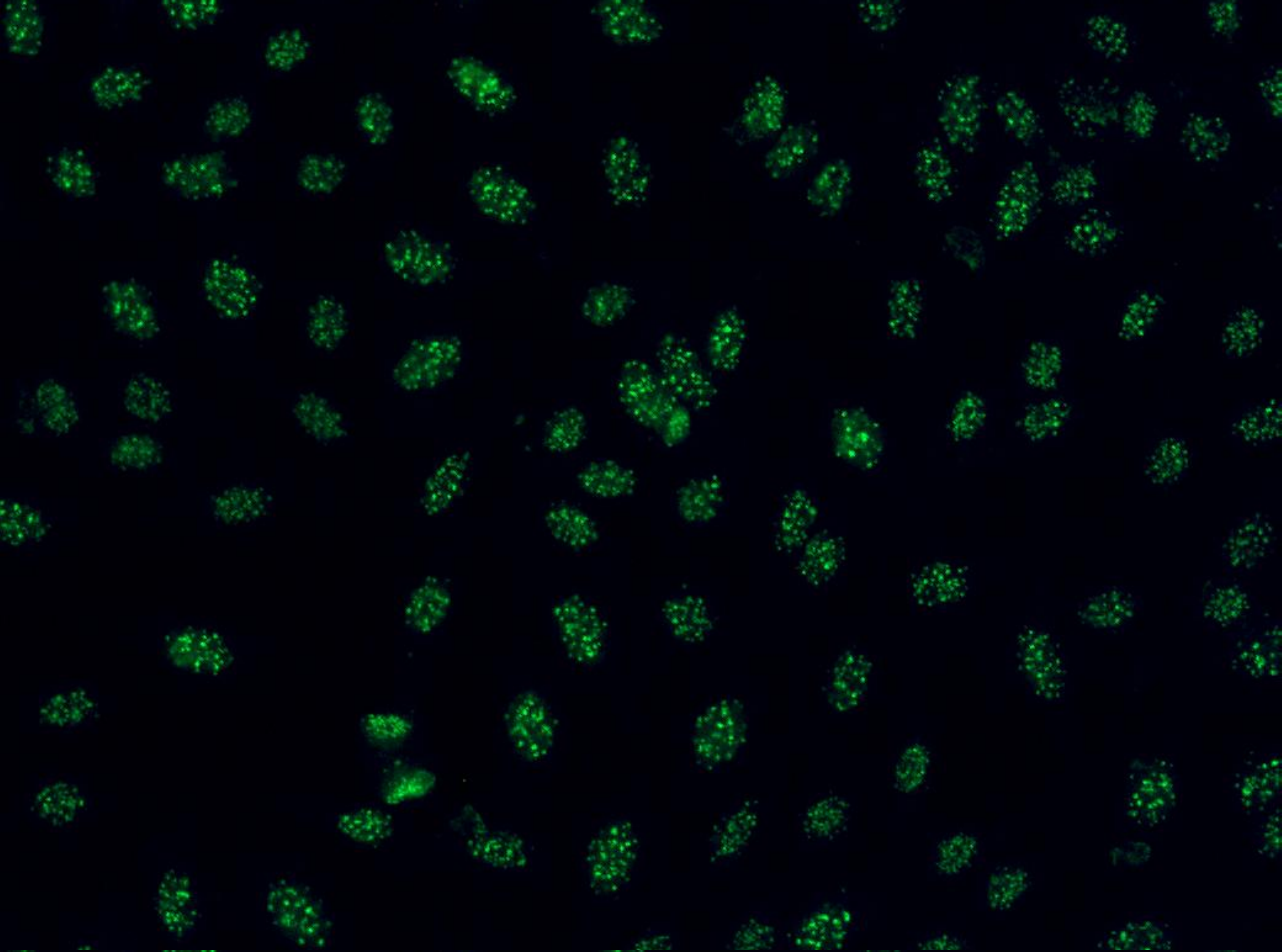}
      {\small centromere}
    \end{minipage}
    \hfill
    \begin{minipage}{0.45\textwidth}
      \centering
      \includegraphics[width=1.0\textwidth]{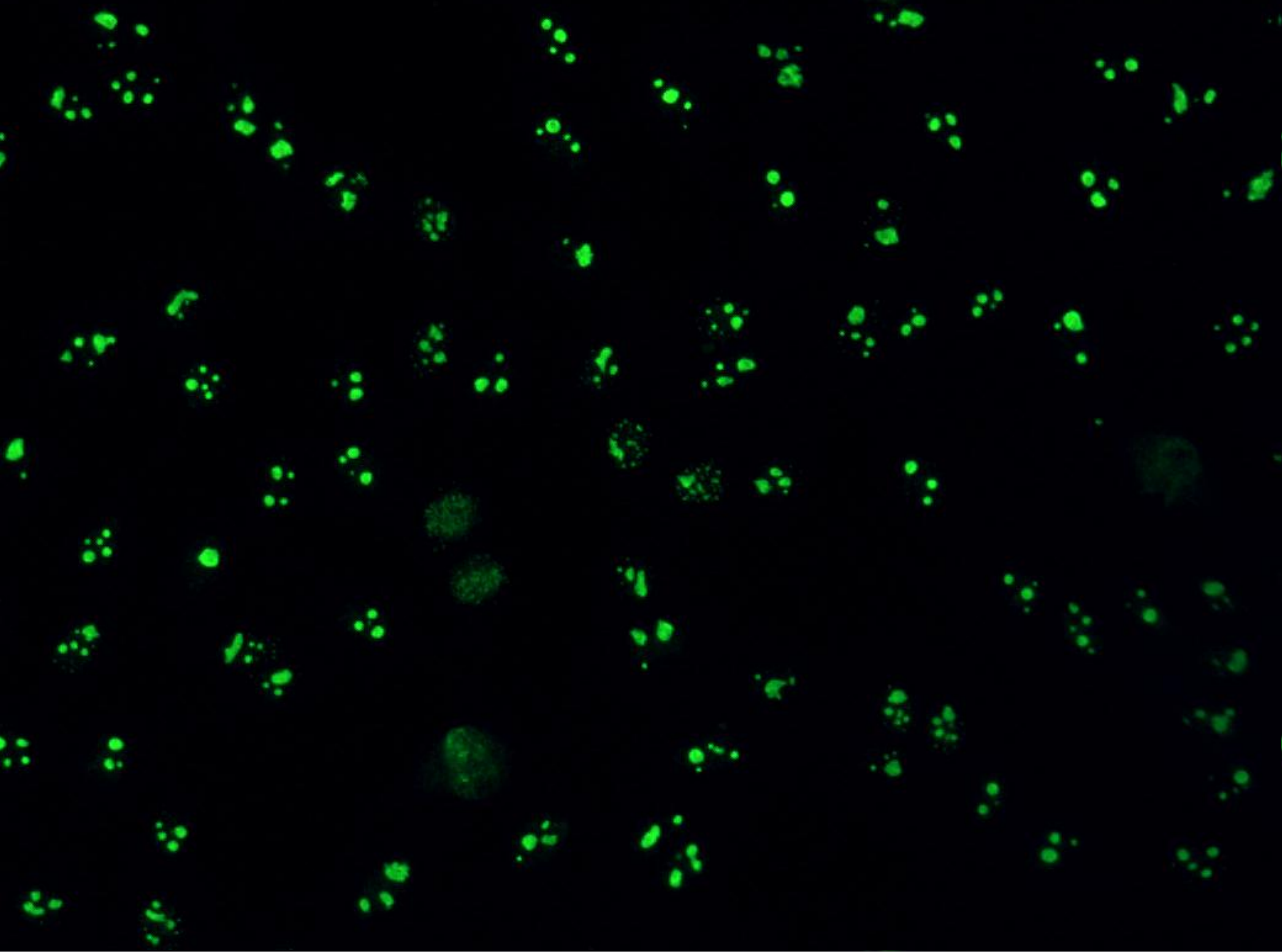}
      {\small nucleolar}
    \end{minipage}
    \hfill
    ~
  \end{minipage}    
  
  ~
  
  \caption{Examples of strong positive ANA specimens.}  
  \label{fig:specimen_images}
\end{figure}

The Anti-Nuclear Antibody (ANA) test is commonly used by clinicians to identify the existence of Connective Tissue
Diseases such as Systemic Lupus Erythematosus, Sjorgren's syndrome, and Rheumatoid Arthritis~\cite{Meroni2010}.
The hallmark protocol for doing this is through Indirect Immunofluorescence (IIF) on Human Epithelial type 2 (HEp-2)
cells~\cite{Meroni2010,Wiik2010}.
This is due to its high sensitivity and the large range expression of antigens. 
Despite its advantages, the IIF method is labour intensive and time consuming~\cite{Bizzaro1998,Pham2005}. 
Each ANA specimen must be examined under a fluorescence microscope by at least two scientists. 
This also renders the test result more subjective, and thus has low reproducibility and large inter-/intra- personnel/laboratory
variabilities~\cite{Hiemann2009,Soda2009}.
To address these issues, it is possible to use Computer Aided Diagnostic (CAD) systems which automatically determine the \mbox{HEp-2} pattern in
the given cell images of a specimen~\cite{Cordelli2011,Elbischger2009,Hiemann2009,Hsieh2009,Perner2002a,Soda2009,Strandmark2012}.
Examples of specimen images are shown in Figure~\ref{fig:specimen_images}.

\begin{table*}
  \small
	\renewcommand{\arraystretch}{1.1}
	\caption
		{
		Existing CAD systems for \mbox{HEp-2} cell classification.
		}
	\label{tab:relatedworks}
	\vspace{0.5ex}
	\centering
	\begin{tabular}{p{3.1cm} | p{7.8cm} | p{5.2cm}}
		\toprule
		\textbf{Approach} 																	& ~\textbf{Descriptors} 																								& ~\textbf{Classifier} \\ 
		\midrule
		Perner~\etal~\cite{Perner2002a} 						& ~Textural 																								& ~Decision Tree \\
		Hiemann~\etal~\cite{Hiemann2009} & ~Structural; textural (1400 features)											& ~LogisticModel Tree\\ 
		Elbischger~\etal~\cite{Elbischger2009} 			& ~Image statistics; cell shape; ~textural ~(9 features) 	  & ~Nearest Neighbour \\																					 
		Hsieh~\etal~\cite{Hsieh2009} 								& ~Image statistics; textural (8 features) 							 	  & ~Learning Vector Quantisation (LVQ) \\
		Soda~\etal~\cite{Soda2009} 									& ~Specific set of features (e.g. textural) for each class & ~Multi Expert System \\ 
 		Cordelli~\etal~\cite{Cordelli2011} 					& ~Image statistics; textural; morphological ~(15 features) & ~AdaBoost \\ 		
 		Strandmark~\etal~\cite{Strandmark2012}                    & ~Morphological; image statistics; textural ~(322 features) & ~Random Forest \\																				 
		\bottomrule
	\end{tabular}
  \vspace{-2ex}
\end{table*}

Properties of existing CAD systems in the literature are shown in Table~\ref{tab:relatedworks}. 
Most of these systems have a common trend: 
they use carefully handpicked features which may only work in a particular laboratory environment and/or microscope configuration.
To address this, several approaches employ a large number of features and apply an automated feature selection process~\cite{Hiemann2009}. 
Another approach uses Multi Expert Systems to allow the use of a specifically tailored feature set and classifier for each \mbox{HEp-2} cell pattern class~\cite{Soda2009}. 
Nevertheless, the generalisation ability of these systems is still not guaranteed since these were only evaluated in one particular dataset with a specific setup. 

One of the most popular approaches for automatic image classification,
here called the codebook approach,
is to express an image in terms of a set of visual words,
selected from a dictionary that has been trained beforehand~\cite{Lazebnik2006,Sanderson2009,Gemert2010}. 
In order to model an image, the codebook approach divides the image into small image patches, 
followed by patch-level feature extraction. 
An encoding process is then employed to compute a histogram of visual words based on these patches. 
Codebook-based descriptors often have higher discrimination power compared to the other image
descriptors~\cite{Lazebnik2006,Gemert2010,Wong_IJCNN_2012}.
Thus, we argue that better classification performance can be achieved by employing such descriptors for CAD systems.

{\bf Contributions}.
In this work we propose the use of a dual-region codebook-based descriptor,
specifically designed to exploit the nature of cell images,
coupled with an adapted form of the Nearest Convex Hull classifier~\cite{Nalbantov2006}.
To our knowledge, this is the first time the codebook approach is applied and studied for the \mbox{HEp-2} cell classification task. 
We evaluate two methods for low-level feature extraction from image patches,
SIFT~\cite{Lowe2004a} and DCT~\cite{Sanderson2009},
in conjunction with three methods for generating the histograms of visual words:
{vector quantisation}~\cite{Gemert2010},
{soft assignment}~\cite{Sanderson2009}
and
{sparse coding}~\cite{Wong_IJCNN_2012}.
We furthermore propose a new \mbox{HEp-2} cell image classification dataset, denoted as \mbox{SNPHEp-2},
which allows the evaluation of the robustness of CAD systems to various hardware configurations.
The number of images is much larger than the existing ICPRContest dataset~\cite{Foggia2010}.

We continue this paper as follows. 
We first delineate the \mbox{HEp-2} cell classification task in Section~\ref{sec:sec_problem}.
In Section~\ref{sec:sec_features} we present the dual-region codebook-based descriptor. 
In Section~\ref{sec:NCH} we overview the Nearest Convex Hull classifier.
Section~\ref{sec:sec_experiment} is devoted to experiments and discussions.
Main findings and future research avenues are given in Section~\ref{sec:sec_conclusions}.

\section{HEp-2 Cell Classification System}
\label{sec:sec_problem}

Each positive \mbox{HEp-2} cell image%
\footnote
  {
  It is assumed that the cell images have been extracted from specimen images 
  via an approach such as background subtraction~\cite{Reddy_TCSVT_2012}.
  }
is represented as a three-tuple {\small$(\Mat{I},\Mat{M},\delta)$} which consists of: 
{\bf (i)}~the Fluorescein Isothiocyanate (FITC) image channel {\small$\Mat{I}$}; 
{\bf (ii)}~a binary cell mask image {\small$\Mat{M}$} which can be manually defined, or extracted from the \mbox{\it 4',6-diamidino-2-phenylindole}
(DAPI) image channel~\cite{Hiemann2009}; and 
{\bf (iii)}~the fluorescence intensity {\small$\delta \in \{\text{strong},\text{weak}\}$} which specifies whether the cell is a strong
positive or weak positive.
Strong positive images normally have more defined details, while weak positive images are duller.

Let {\small$\Mat{Y}$} be a probe image {\small$\Mat{Y} = (\Mat{I},\Mat{M},\delta)$}, and $\ell$ be its class label. 
Given a gallery set
{\small $\mathcal{G} = \{ (\Mat{I},\Mat{M},\delta)^{\mathcal{G}}_1, (\Mat{I},\Mat{M},\delta)^{\mathcal{G}}_2, \ldots, (\Mat{I},\Mat{M},\delta)^{\mathcal{G}}_m\}$},
the task of a classifier {\small$\varphi:\Mat{Y} \times \mathcal{G} \mapsto \widehat{\ell}$} is to produce {\small$\widehat{\ell}$},
where ideally {\small$\widehat{\ell} = \ell$}. 

\begin{figure}[!b]
  \centering
  \includegraphics[width=1.0\linewidth]{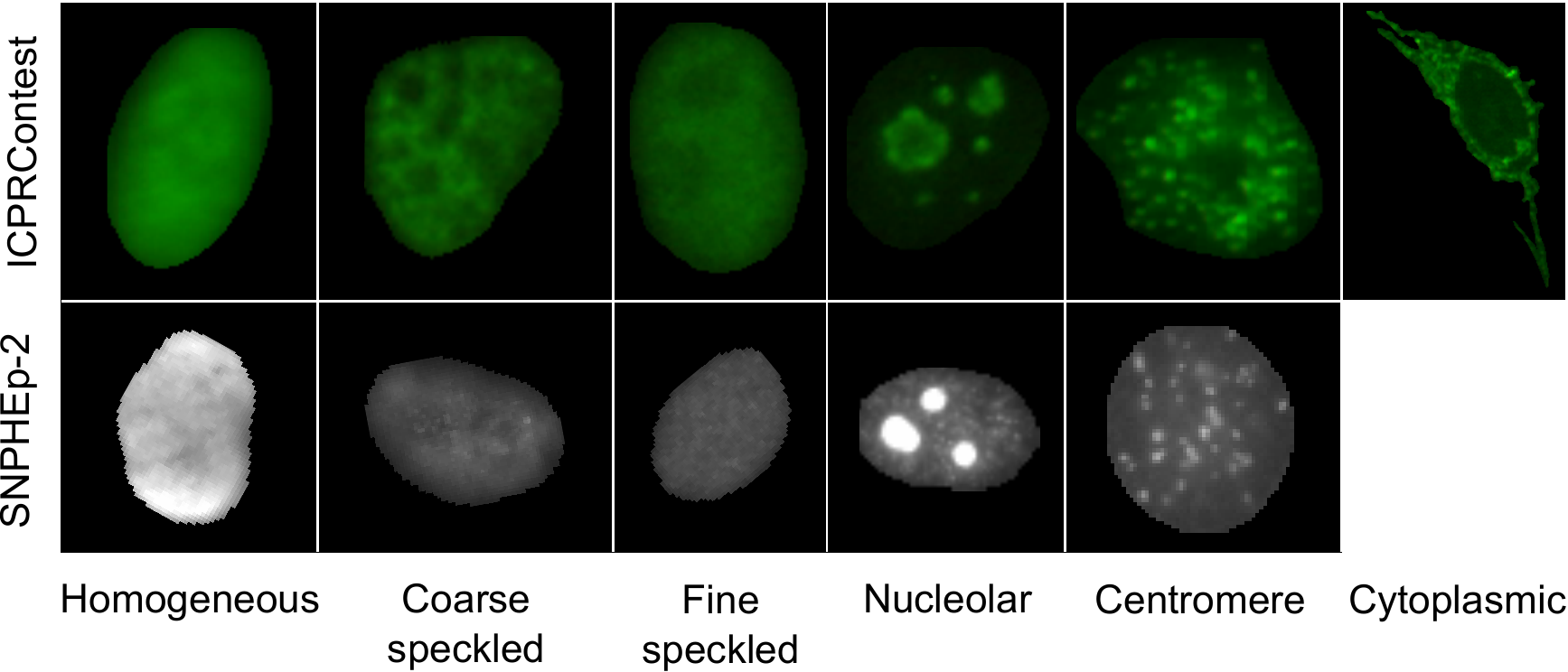}
  \caption{Sample images from ICPRContest dataset~\cite{Foggia2010} and the proposed \mbox{SNPHEp-2} dataset.}
  \label{fig:dataset} 
\end{figure}

We consider six \mbox{HEp-2} cell patterns~\cite{Wiik2010} listed below;
example images are shown in Fig.~\ref{fig:dataset}.

\begin{small}
\begin{enumerate}[{\bf (1)}]
\renewcommand{\itemsep}{0ex}

\item
{\it homogeneous}: a uniform diffuse fluorescence covering the entire nucleoplasm sometimes accentuated in the nuclear periphery

\item
{\it coarse speckled}: densely distributed, variously sized speckles, generally associated with larger speckles, throughout nucleoplasm of interphase cells;
 nucleoli are negative

\item
{\it fine speckled}: fine speckled staining in a uniform distribution, sometimes very dense so that an almost homogeneous pattern is attained; 
 nucloli may be positive or negative

\item
{\it nucleolar}: brightly clustered larger granules corresponding to decoration of the fibrillar centers of the nucleoli as well as the coiled bodies

\item
{\it centromere}: rather uniform discrete speckles located throughout the entire nucleus

\item
{\it cytoplasmic}: a very fine dense granular to homogeneous staining or cloudy pattern covering part or the whole cytoplasm
\end{enumerate}
\end{small}

\renewcommand{\baselinestretch}{0.98}\small\normalsize
\section{HEp-2 Cell Image Descriptor}
\label{sec:sec_features}

\begin{figure}[!b]
	\centering
	\includegraphics[width=1\columnwidth]{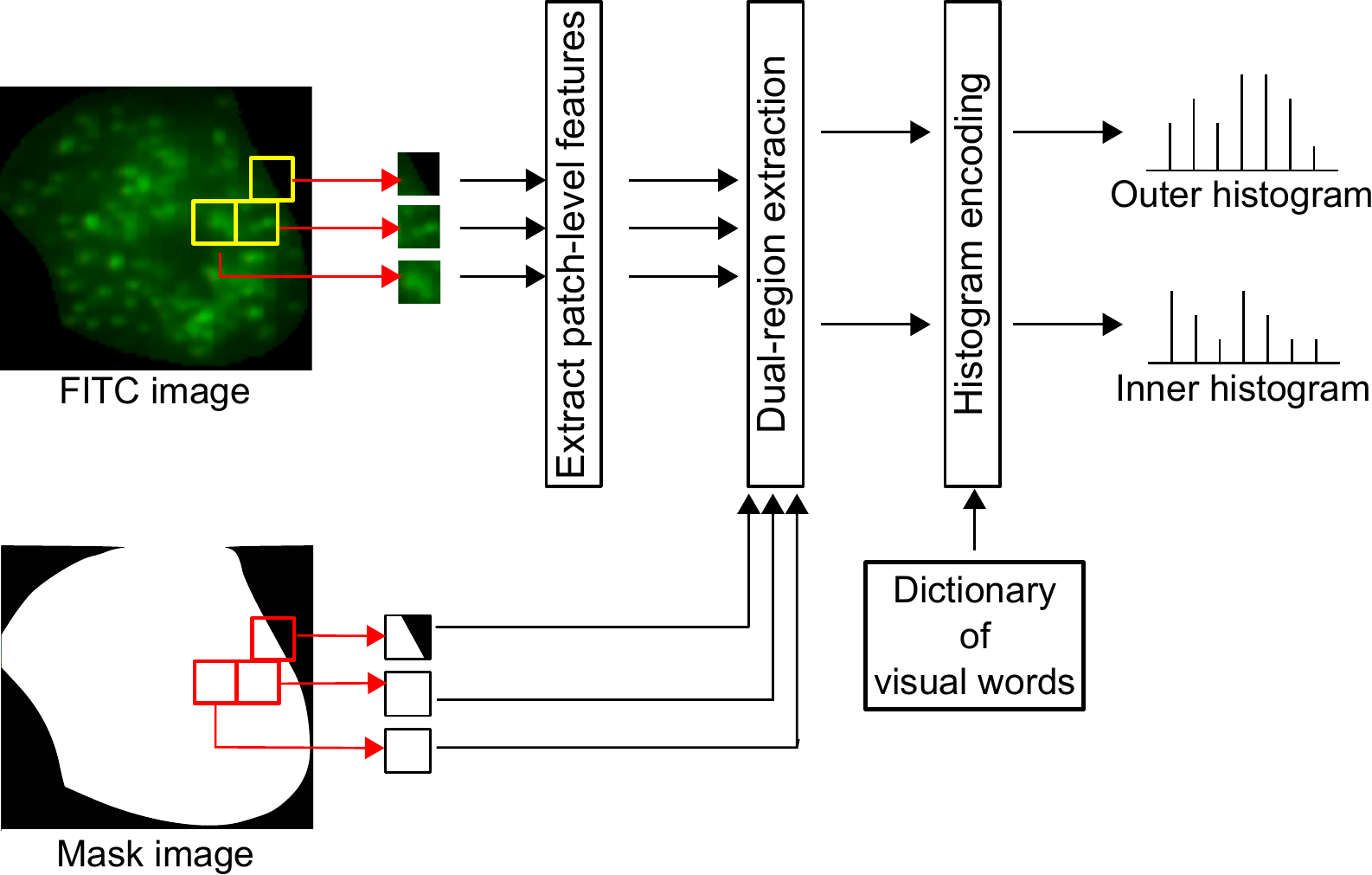}	
	\caption
	  {
	  Conceptual diagram for the proposed HEp-2 cell image descriptor. 
	  Both the FITC image and its corresponding mask image are divided into small overlapping patches. 	
	  Patch-level features are extracted from FITC patches. 
	  Each FITC patch is then classified into either outer or inner region by using information extracted from its corresponding mask patch. 
	 Finally, both inner and outer histograms are obtained by an encoder employing a learned dictionary of visual words.
	  } 
	\label{fig:system_diagram}	
\end{figure}

The overall idea of the proposed HEp-2 cell image descriptor is shown in Fig.~\ref{fig:system_diagram}. 
Each cell is divided into small overlapping patches.
The patches are then used to construct two histograms of visual words:
{\it inner} and {\it outer} histograms,
depending on whether the patches come from the the inside of the cell, or its edges, respectively.
We first describe low-level patch-level features in Section~\ref{sec:patch_level_feature_extraction},
followed by presenting the dual-region structure in Section~\ref{sec:dual_region_structure}.
In Section~\ref{sec:histogram_generation} we present several histogram encoding methods.

%
%
%

\subsection{Patch-level Feature Extraction}
\label{sec:patch_level_feature_extraction}

Given a HEp-2 cell image {\small$(\Mat{I},\Mat{M},\delta)$},
both the FITC image $\Mat{I}$ and mask image $\Mat{M}$ are divided into small overlapping patches
{\small $\mathcal{P}_I = \{\Mat{p}_{I,1},~ \Mat{p}_{I,2},~ \ldots,~ \Mat{p}_{I,n}\}$}
and
{\small $\mathcal{P}_M = \{\Mat{p}_{M,1},~ \Mat{p}_{M,2},~ \ldots,~ \Mat{p}_{M,n}\}$}.
The division is accomplished in the same manner of both images,
resulting in each patch in the FITC image having a corresponding patch in the mask image.
Let {\small $\Vec{f}$} be a patch-level feature extraction function
{\small $\Vec{f}: \Mat{p}_{I} \mapsto \Vec{x}$}, where {\small $\Vec{x} \in \mathbb{R}^d$}. 
{\small $\mathcal{P}_{I}$} now can be represented as {\small $\mathcal{X}=\{\Vec{x}_1,~ \Vec{x}_2,~ \ldots,~ \Vec{x}_n\}$}.

We selected two patch-level feature extraction techniques, based on
the Scale Invariant Feature Transform (SIFT)
and the Discrete Cosine Transform (DCT). 
The low-level SIFT descriptor is invariant to uniform scaling, orientation
and partially invariant to affine distortion and illumination changes~\cite{Lowe2004a}.
These attributes are advantageous in this classification task as cell images are unaligned and have high within class variabilities. 
DCT based features proved to be effective for face recognition in video surveillance~\cite{Sanderson2009,Wong_IJCNN_2012}.
By using only the low frequency DCT coefficients (essentially a low-pass filter),
each patch representation relatively robust to small alterations~\cite{Sanderson2009}. 
We follow the extraction procedures for SIFT and DCT as per~\cite{Liu2011} and~\cite{Sanderson2009}, respectively.

%
%
%

\subsection{Dual-Region Structure}
\label{sec:dual_region_structure}

\begin{figure}[!b]
  \centering
  \includegraphics[width=0.7\linewidth]{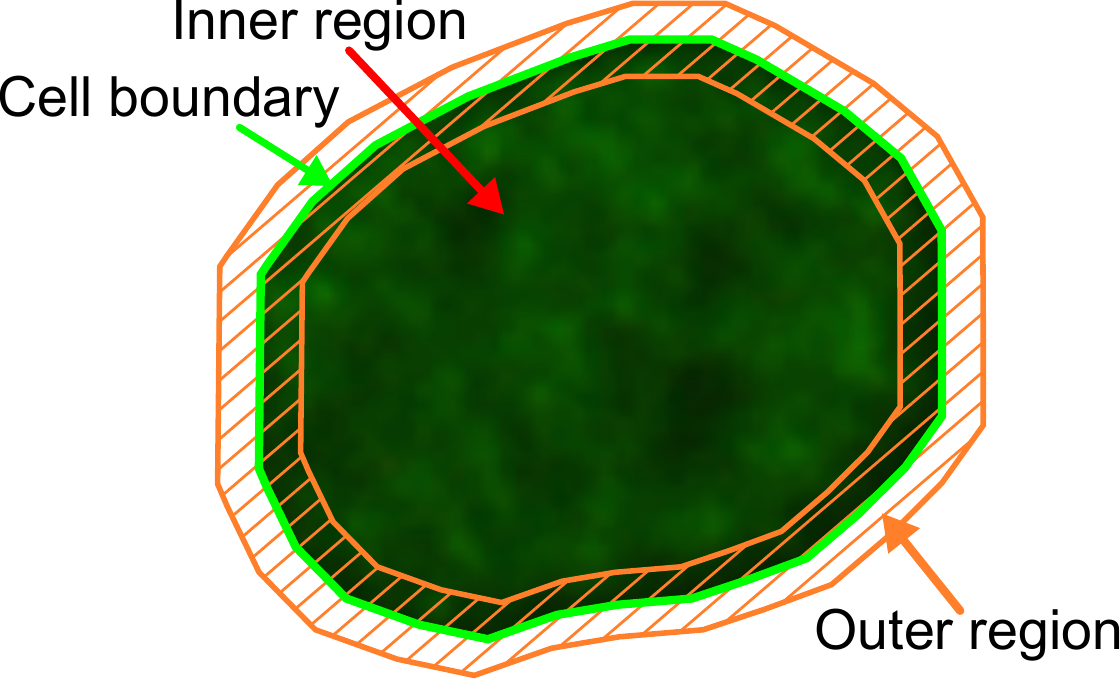}  
  \caption{
  Conceptual diagram for the proposed dual-region structure.
  The green line is the cell boundary.  
  The outer region is denoted by the striped patterns, and inner region is the area denoted by the red arrow.
  }
  \label{fig:2region} 
\end{figure}

We aim to model cell characteristics by building separate histograms for the inner region
(which is often a uniform texture) and the outer region (which contains information related to cell edges and shape).
To this end, each FITC patch is first classified as either belonging to the inner or outer region by inspecting its corresponding mask patch. 
Fig.~\ref{fig:2region} shows how the regions are imposed on a cell image.

Let {\small $\mathcal{X} = \mathcal{X}_o \cup \mathcal{X}_i$},
with {\small $\mathcal{X}_o$} representing the set of outer patches,
and {\small $\mathcal{X}_i$} the set of inner patches.
The classification of patch {\small $\Mat{p}_{I}$} into a region is done via:

\vspace{-1ex}
\begin{small}
\begin{equation}
	\Mat{p}_{I} \in  \left\{
	\begin{array}{l l}	
	\mathcal{X}_o & \quad \text{if} ~ \tau_1 ~ \leq ~ \operatorname{fg}(\Mat{p}_{M}) ~ < ~ \tau_2 \\ 
	\mathcal{X}_i & \quad \text{if} ~ \tau_2 ~ \leq ~ \operatorname{fg}(\Mat{p}_{M})\\	
	\end{array}
	\right.	
\end{equation}
\end{small}

\vspace{-1ex}
\noindent
where {\small $\Mat{p}_{M}$} is the corresponding mask patch;
{\small $\operatorname{fg}(\Mat{p}_{M})$} computes the percentage of foreground pixels from mask patch {\small $\Mat{p}_{M}$};
$\tau_1$ is the minimum foreground pixel percentage of a patch belonging to the outer region;
and $\tau_2$ is the maximum foreground pixel percentage of a patch belonging to the outer region,
as well as the minimum pixel percentage of a patch belonging to the inner region.

%
%
%

\subsection{Generation of Histograms}
\label{sec:histogram_generation}

Let {\small $\mathcal{X}_r$} be the set of patch-level features for either the inner or outer region
(ie.,~{\small $\mathcal{X}_r = \mathcal{X}_i$} or {\small $\mathcal{X}_r = \mathcal{X}_o$}).
For each patch-level feature {\small $\Vec{x}_j \in \mathcal{X}_r$},
a local histogram {\small $\Vec{h}_j$} is obtained by an encoding method.
The overall histogram representation for region $r$ is then obtained via averaging~\cite{Sanderson2009,Wong_IJCNN_2012}:

\vspace{-1ex}
\noindent
\begin{small}
\begin{equation}
  \Vec{H}^{[r]} = \frac{1}{|\mathcal{X}_r|} \sum\nolimits_{j=1}^{|\mathcal{X}_r|} \Vec{h}_j
\end{equation}
\end{small}

\vspace{-1ex}
\noindent
where {\small $|\mathcal{X}_r|$} is the number of elements in set {\small $\mathcal{X}_r$}.
In this work we consider three popular histogram encoding methods:
(1)~vector quantisation; (2)~soft assignment; (3)~sparse coding. 
The methods are elucidated below.

\subsubsection{Vector Quantisation (VQ)}

Given set {\small $\mathcal{D}$}, a dictionary of visual words,
the $i$-th dimension of local histogram {\small $\Vec{h}_j$} for patch {\small $\Vec{x}_j$} is computed via:

\vspace{-1ex}
\begin{small}
\begin{equation}
	\Vec{h}_{j,i} = \left\{
	\begin{array}{l l}
	1 & \quad \text{if } i = \argmin{k \in 1, \ldots, |\mathcal{D}|}  {\phi(k) = \operatorname{dist}(\Vec{x}_j,\Vec{d}_k)} \\
	0 & \quad \text{otherwise}
	\end{array}
	\right.
	\label{eqn:vq_encoding}
\end{equation}
\end{small}

\vspace{-1ex}
\noindent 
where {\small $\operatorname{dist}(\Vec{x}_j,\Vec{d}_k)$}
is a distance function between {\small $\Vec{x}_j$} and {\small $\Vec{d}_k$},
with {\small $\Vec{d}_k$} the {\small $k$}-th entry in the dictionary {\small $\mathcal{D}$}.
The dictionary is obtained via the $k$-means algorithm~\cite{Bishop2006} on training patches.

The VQ approach is considered as a hard assignment approach since each image patch is only assigned to one of the visual words.
This hard assignment is sensitive to noise~\cite{Gemert2010}.

\subsubsection{Soft Assignment (SA)}

In comparison to the VQ approach above,
a more robust approach is to apply a probabilistic method~\cite{Sanderson2009}.
Here the visual dictionary {\small $\mathcal{D}$} is a convex mixture of Gaussians.
The $i$-th dimension of the local histogram for {\small $\Vec{x}_j$} is calculated as: 

\vspace{-1ex}
\begin{small}
\begin{equation}
	\Vec{h}_{j,i} = \frac{w_i p_i(\Vec{x}_j)}{\sum_{k=1}^{|\mathcal{D}|}{ w_k ~ p_k( \Vec{x}_j ) }}
	\label{eqn:probabilistic_encoding}
\end{equation}
\end{small}

\vspace{-1ex}
\noindent
where {\small $p_i( \Vec{x} )$} is the likelihood of {\small$\Vec{x}$} according to the $i$-th component of the visual dictionary:

\vspace{-1ex}
\begin{small}
\begin{equation}
  p_i(\Vec{x})  = \frac
    {
    \exp \left[ -\frac{1}{2}\left(  \Vec{x} - \Vec{\mu}_i \right)^T \Mat{C}_i^{-1} \left(
    \Vec{x} - \Vec{\mu}_i \right) \right] }
    {
    \left( 2 \pi \right)^\frac{d}{2} | \Mat{C}_i | ^\frac{1}{2}
    }
\end{equation}
\end{small}

\vspace{-1ex}
\noindent
with {\small $w_i$}, {\small $\Vec{\mu}_i$} and {\small$\Mat{C}_i$}
representing the weight, mean vector and covariance matrix of Gaussian $i$, respectively.
The scalar $d$ represents the dimensionality of {\small $\Vec{x}$}.
The dictionary {\small $\mathcal{D}$} is obtained using the Expectation Maximisation algorithm~\cite{Bishop2006} on training patches.

\subsubsection{Sparse Coding (SC)}

It has been observed that each local histogram produced via Eqn.~(\ref{eqn:probabilistic_encoding})
is sparse in nature (ie., most elements are close to zero)~\cite{Wong_IJCNN_2012}.
As such, it is possible to adapt dedicated sparse coding algorithms
in order to represent each patch as a combination of dictionary atoms~\cite{Coates2011,Yang2009}. 

A vector of weights {\small$\Vec{\alpha} = \left[ \alpha_1, \alpha_2,...,\alpha_n \right]^T$}
is computed for each {\small$\Vec{x}_j$} by solving a minimisation problem that selects a sparse set of dictionary atoms. 
As the theoretical optimality of the \mbox{$\ell_1$-norm} minimisation solution is guaranteed~\cite{Tropp2010},
in this work we used:

\vspace{-1ex}
\begin{small}
\begin{equation}
	\operatorname{min}\frac{1}{2}\|\Mat{D}\Vec{\alpha}-\Vec{x}_j\|^2_2 + \lambda \sum\nolimits_k{\|\alpha_k\|_1}			
	\label{eq:SCObjective}
\end{equation}
\end{small}

\vspace{-1ex}
\noindent
where
{\small $\| \cdot \|_p$} denotes the \mbox{\small $\ell_p$-norm}
and
{\small $\Mat{D} \in \mathbb{R}^{d \times n}$} is a matrix of dictionary atoms.
The dictionary {\small$\Mat{D}$} is trained by using the K-SVD algorithm~\cite{Aharon2006}.

As $\Vec{\alpha}$ can have negative values due to the objective function in Eqn.~(\ref{eq:SCObjective}),
we construct each local histogram using the absolute value of each element in {\small $\Vec{\alpha}$}~\cite{Wong_IJCNN_2012}:

\vspace{-1ex}
\begin{small}
\begin{equation}
  \Vec{h}_j = \left[~ |\alpha_1|,~ |\alpha_2|,~ \ldots,~ |\alpha_n|~ \right]
\end{equation}
\end{small}

\vspace{-1ex}
Compared to both Eqns.~(\ref{eqn:vq_encoding}) and~(\ref{eqn:probabilistic_encoding}),
obtaining the histogram using sparse coding is considerably more computationally intensive,
due to the need to solve a minimisation problem for each patch.

\renewcommand{\baselinestretch}{0.95}\small\normalsize
\section{Classifiers}
\label{sec:NCH}

Let set {\small ${Q}_{X} = \left\{ \Mat{H}^{[i]}_{X},~ \Mat{H}^{[o]}_{X} \right\}$}
represent the average inner and outer histograms for cell image {\small $X$}.
Below we describe two classifiers that we have adapted to use both the inner and outer regions:
(1)~nearest neighbour (NN),
and
(2)~nearest convex hull (NCH).

\subsection{Nearest Neighbour (NN)}
\label{sec:nearest_neighbour_classifier}

The NN classifier assigns the class of probe image {\small $A$} to be the class of the nearest training image {\small $B$}.
For the purposes of this classifier, 
and inspired by~\cite{Tommasi2008},
we define the distance between images {\small $A$} and {\small $B$} as:

\vspace{-1ex}
\begin{small}
\begin{equation}
	\operatorname{d}  \left( Q_{A},Q_{B} \right)
	\mbox{=~}
	\gamma  
	\left\| \Mat{H}^{[i]}_{A} \mbox{~-~} \Mat{H}^{[i]}_{B} \right\|_p
	+
	(1 \mbox{~-~} \gamma)
  \left\| \Mat{H}^{[o]}_{A} \mbox{~-~} \Mat{H}^{[o]}_{B} \right\|_p
	\label{eq:distDR}
\end{equation}
\end{small}

\vspace{-1ex}
\noindent
where $\gamma \in [0,1]$ is a mixing parameter found during training.

\subsection{Nearest Convex Hull (NCH)}
\label{sec:convex_hull_classifier}

In NCH, each training class is approximated with a simple convex model,
or more specifically, the convex hull of the descriptors of the training images~\cite{Nalbantov2006}.
This reduces the sensitivity to within class variation,
as ``missing'' samples can be approximated using the convex model~\cite{Cevikalp2010}. 

Let {\small $\lambda_{C} = \left\{ \Omega^{[i]}_{C}, \Omega^{[o]}_{C} \right\} $} denote the model of class {\small $C$},
comprised of inner and outer components.
In order to take into account both the inner and outer histograms,
we define distance between image~{\small $A$} and class~{\small $C$} as:

\vspace{-1ex}
\begin{small}
\begin{equation}
  \operatorname{d}  \left( Q_{A}, \lambda_{C} \right)
  \mbox{=~}
  \gamma
  \operatorname{d}_\mathtt{NCH} \hspace{-0.25ex} \left(\hspace{-0.25ex} \Mat{H}^{[i]}_{A} , \Omega^{[i]}_{C} \right)
  \hspace{-0.25ex}\mbox{+}\hspace{0.25ex}
  (1 \mbox{-~}\gamma)
  \operatorname{d}_\mathtt{NCH} \hspace{-0.25ex} \left(\hspace{-0.25ex} \Mat{H}^{[o]}_{A} , \Omega^{[o]}_{C} \right)
  \label{eq:distNCH}
\end{equation}
\end{small}

\vspace{-1ex}
\noindent
where {\small $\gamma \in [0,1]$} is a mixing parameter found during training,
and {\small $\operatorname{d}_{NCH}(\Mat{H}, \Omega_{C})$}
is the distance between histogram {\small $\Mat{H}$} and convex model {\small $\Omega_{C}$}, defined as:

\vspace{-1ex}
\begin{small}
\begin{equation}
	\operatorname{d}_{NCH} \left( \Mat{H}, \Omega_{C} \right) = \operatorname{min} \| \Mat{H} - \Vec{\omega} \|_1, ~ \Vec{\omega} \in \Omega_{C}
\end{equation}
\end{small}

\noindent
where {\small $\Omega_C$} is a set of points generated by a linear combination of the training samples. 
Given a set of training histograms for class {\small $C$},
{\small $\{ \Mat{H}_1,~ \Mat{H}_2,~ \ldots,~ \Vec{H}_m \}$},
each member of {\small $\Omega_C$} is defined as~\cite{Nalbantov2006}:

\vspace{-1ex}
\begin{small}
\begin{equation}
	\Vec{\omega} = \sum\nolimits_{i=1}^{m}{\beta_i \Mat{H}_i}, \text{~~subject to~~} \sum\nolimits_{i=1}^{m}{\beta_i} = 1
\end{equation}
\end{small}

\vspace{-1ex}
The above model implicitly treats any combination of histogram descriptors as a valid gallery example.

\renewcommand{\baselinestretch}{1.00}\small\normalsize
\section{Experiments}
\label{sec:sec_experiment}

In this section we first overview the two datasets used in the experiments.
We then evaluate the six variants of the codebook-based descriptor,
where each of two low-level feature extraction techniques (SIFT and DCT)
is coupled with three possible methods for generating the histograms of visual words
(VQ, SA, and SC).
Finally we compare the best codebook-based variant against two recently proposed systems.
The various systems were implemented with the aid of the Armadillo C++ library~\cite{Armadillo_2010}.

\subsection{ICPR \mbox{HEp-2} Contest Dataset}
\label{sec:icprcontest_dataset}

The ICPR \mbox{HEp-2} Cell Classification Contest Dataset (ICPRContest)~\cite{Foggia2010}
contain 1,457 cells extracted from 28 specimen images.
It contains six patterns: centromere, coarse speckled, cytoplasmic, fine speckled, homogeneous, and nucleolar. 
Each specimen image was acquired by means of fluorescence microscope (40-fold magnification) coupled with 50W mercury vapour lamp and with a CCD camera. 
The cell image masks were hand labelled. 
See Fig.~\ref{fig:dataset} for examples. 

As the official test set is not yet publicly available,
we use only the official training set to create ten-fold validation sets.
The available images are divided into training and testing sets with 14 specimens each. 
As such, in each set, each pattern class only has 1-2 specimen images,
where a specimen image contains a set of cells having the same pattern
(see Fig.~\ref{fig:specimen_images} for a visual representation). 
In total there are 721 and 736 cell images extracted for training and testing, respectively. 
The validation sets were created by randomly selecting the images from the 721 cell images. 
Each fold contains 652 and 69 cell images for training and testing respectively.

Note that due to the abovementioned limitation of available images,
each class can only have 1-2 specimen images.
As such there is an assumed bias, as there is a high chance that cells extracted from the same specimen image
exist in both the training and testing sets of each fold.

\subsection{SNP \mbox{HEp-2} Cell Dataset}

The SNP \mbox{HEp-2} Cell Dataset (SNPHEp-2)
was obtained between January and February 2012 at Sullivan Nicolaides Pathology laboratory, Australia. 
The dataset\footnote{The SNPHEp-2 dataset is available for download at \mbox{\url{http://itee.uq.edu.au/~lovell/snphep2/}}}
has five patterns: centromere, coarse speckled, fine speckled, homogeneous and nucleolar. 
The 18-well slide of HEP-2000 IIF assay from Immuno Concepts N.A.~Ltd.~with screening dilution 1:80 was used to prepare 40 specimens. 
Each specimen image was captured using a monochrome high dynamic range cooled microscopy camera,
which was fitted on a microscope with a plan-Apochromat 20x/0.8 objective lenses and an LED illumination source. 
DAPI image channel was used to automatically extract the cell image masks. 

There are 1,884 cell images extracted from 40 specimen images. 
The specimen images are divided into training and testing sets with 20 images each (4 images for each pattern). 
In total there are 905 and 979 cell images extracted for training and testing.
Five-fold validation of training and testing were created by randomly selecting the training and test images. 
Both training and testing in each fold contain around 900 cell images (approx. 450 images each).
Examples are shown in Fig.~\ref{fig:dataset}. 

By using the SNPHEp-2 dataset in addition to the ICPRContest dataset, we obtain the following benefits:
(1)~the specimens of both datasets were not captured by the same microscope configuration
(eg.,~the microscope's objective lens magnitude for ICPRContest is 40x, while 20x for \mbox{SNPHEp-2}),
allowing us to test the robustness of CAD systems to variations in image capture conditions;
(2)~there is no bias compared to the ICPRContest experiment setup,
allowing for a more thorough evaluation.

\subsection{Codebook-Based Descriptor Variants}
\label{sec:exp_descriptor_variants}

In this section we evaluate the discriminative power of the codebook-based descriptor,
with and without the dual-region structure. 
When the dual-region structure is not employed,
each cell image is represented by one histogram constructed from both the inner and outer patches.

As there are three histogram encoding methods
(ie.,~VQ, SA and SC) and two patch-level features (ie.,~SIFT and DCT),
there are six variants of the codebook-based descriptor. 
For clarity, each variant is styled as:
\textit{[patch-level features]}-\textit{[histogram encoding method]}.
For example, the variant using DCT as its patch-level features and VQ as its encoding method is called DCT-VQ. 

The NN classifier was employed in this comparison,
in order to reduce the total number of combinations.
Based on preliminary experiments, we selected \mbox{$\ell_1$-norm} distance in Eqn.~(\ref{eq:distDR})
to measure the distance between two images. 
All other hyperparameters of each approach were found in the training set of each cross-validation set.

Table~\ref{tab:2region} presents the average Correct Classification Rate (CCR)
for each descriptor variant on the ICPRContest and SNPEHEp-2 datasets,
using both single- and dual-region configurations.
We can observe that all variants have higher CCR on ICPRContest than on SNPEHEp-2,
which is consistent with the bias in the ICPRContest dataset setup. 

\begin{table}[!t]
  \centering
  \footnotesize
  \caption
    {
    Performance comparison of codebook-based descriptor variants on the ICPRContest and SNPHEp-2 datasets,
    using the NN classifier.
    The scores are shown as average correct classification rate (in percentage) along with their standard deviations.
    SR = single region; DR = dual region.
    }
  \label{tab:2region}
  \vspace{0.5ex}
  \begin{tabular}{l|cc|cc}
    \toprule
    \hspace{-1.5ex}\textbf{Descriptor} & \multicolumn{2}{c|}{\textbf{ICPRContest}} & \multicolumn{2}{c}{\textbf{SNPHEp-2}} \\               
    \hspace{-1.5ex}\textbf{Variant}    & SR & DR  & SR & DR  \\ 
    \midrule
    \hspace{-1.5ex}DCT-SA  & \textbf{93.8~$\pm$~2.2}  & \textbf{94.9~$\pm$~2.1} & \textbf{74.7~$\pm$~3.6} & \textbf{76.7~$\pm$~2.5}\hspace{-1.5ex} \\ 
    \hspace{-1.5ex}DCT-VQ        & 89.5~$\pm$~3.4  & 90.6~$\pm$~3.2  & 72.3~$\pm$~3.2 & 73.6~$\pm$~2.1\hspace{-1.5ex}  \\ 
    \hspace{-1.5ex}DCT-SC   & 81.7~$\pm$~3.4  & 84.8~$\pm$~3.5  & 59.9~$\pm$~2.7 & 63.6~$\pm$~1.8\hspace{-1.5ex}  \\ 
    \midrule 
    \hspace{-1.5ex}SIFT-SA  & 80.1~$\pm$~3.8  & 82.6~$\pm$~3.5 & 56.6~$\pm$~3.1 & 59.2~$\pm$~2.5\hspace{-1.5ex} \\ 
    \hspace{-1.5ex}SIFT-VQ       & 86.4~$\pm$~4.0  & 86.8~$\pm$~3.7  & 64.7~$\pm$~2.3 & 64.9~$\pm$~2.5\hspace{-1.5ex}  \\ 
    \hspace{-1.5ex}SIFT-SC & 79.9~$\pm$~3.3  & 86.1~$\pm$~2.6  & 66.4~$\pm$~2.9 & 67.9~$\pm$~2.5\hspace{-1.5ex}  \\ 
    \bottomrule
  \end{tabular}
\end{table}

The DCT-SA variant is more discriminative and robust to various hardware configurations,
as it consistently outperforms the other variants on both datasets. 
This high performance can be partly attributed to effect of soft-assignment,
which can be more expressive than the other variants~\cite{Gemert2010}.

Generally, DCT has better performance than SIFT on most codebook-based variants on both datasets.
The only exception is on \mbox{SNPHEp-2}, where the SIFT-SC outperforms DCT-SC. 
This suggests that for this application, DCT is more suitable than SIFT for representing low-level patch features.

The results also show that imposing a spatial structure (ie.,~using the dual-region setup in contrast to the single-region setup)
increases the performance of all of the variants,
while generally reducing the standard deviation.

\subsection{Classifier Variants}
\label{sec:exp_classifier_variants}

In this section  we compare the performance of the NN and NCH classifiers on both datasets. 
We use the most discriminative descriptor found in Section~\ref{sec:exp_descriptor_variants}, 
ie.,~DCT-SA.
In addition, we have also evaluated the performance of two baseline descriptors:
(i) raw image, where a raw cell image is vectorised,
and
(ii) rotation invariant Linear Binary Patterns (LBP)~\cite{Ojala2002}, using a configuration of 8 neighbours and 1 pixel radius.
For the baseline descriptors, the cell images were used without any further processing (eg.,~no spatial structure was imposed).

The results, presented in Table~\ref{tab:NCH_comparisons},
show that NCH generally outperforms NN regardless of the descriptor being used.
An exception is LBP on SNPHEp-2, where NN performs slightly better.
In most cases the NCH classifier also provides the most improvement on the more difficult SNPHEp-2 dataset.
with its performance on the ICPRContest dataset close to the performance of NN.
The results also show that the proposed DCT-SA approach (in both single- and dual-region configurations)
considerably outperforms LBP, especially on the SNPHEp-2 dataset.

\begin{table}[!t]
  \centering
  \footnotesize
  \caption
    {
    Performance comparison of the NN and NCH classifiers on the ICPRContest and SNPHEp-2 datasets.
    SR~=~single region; DR~=~dual region.
    }
  \label{tab:NCH_comparisons}
  \vspace{0.5ex}
  \begin{tabular}{l|cc|cc}
    \toprule
    \hspace{-1.5ex}\textbf{Descriptor} & \multicolumn{2}{c|}{\textbf{ICPRContest}} & \multicolumn{2}{c}{\textbf{SNPHEp-2}}\hspace{-1.5ex}\\                             
    & NN & NCH & NN & NCH  \\ 
    \midrule
    \hspace{-1.5ex}DCT-SA + DR  & 94.9~$\pm$~2.1  & {\bf 95.5~$\pm$~2.2}  & 76.7~$\pm$~2.5 & {\bf 80.6~$\pm$~2.1}\hspace{-1.5ex}  \\    
    \hspace{-1.5ex}DCT-SA + SR  & 93.8~$\pm$~2.2  & 94.3~$\pm$~2.3  & 74.7~$\pm$~3.6 & 78.5~$\pm$~3.2\hspace{-1.5ex}  \\ 
    \midrule 
    \hspace{-1.5ex}LBP          & 85.8~$\pm$~2.6  & 86.4~$\pm$~3.1 & 49.9~$\pm$~3.8 & 47.6~$\pm$~2.2\hspace{-1.5ex}  \\ 
    \hspace{-1.5ex}Raw Image    & 39.8~$\pm$~3.2  & 57.7~$\pm$~4.5 & 39.1~$\pm$~3.3 & 43.1~$\pm$~4.2\hspace{-1.5ex}  \\         
    \bottomrule
  \end{tabular}
\end{table}

\subsection{Comparative Evaluation of Systems}

In this section we compare the best performing codebook-based system found in Section~\ref{sec:exp_classifier_variants},
ie.,~dual-region DCT-SA coupled with the NCH classifier,
against two recently proposed systems in Cordelli~et~al.~\cite{Cordelli2011} and Strandmark~et~al.~\cite{Strandmark2012}.

We implemented the best reported descriptor in~\cite{Cordelli2011},
which is comprised of features such as image energy, mean and entropy, calculated from intensity and LBP channels.
The LBP channel is computed by computing the local pattern code for each pixel in the intensity channel.
We selected Logistic Boosting (LogitBoost) as the classifier instead of AdaBoost as the former obtained better performance.
We denote this system as {\it Cordelli}.

We denote the system in~\cite{Strandmark2012} as {\it Strandmark},
and used the code provided by the authors. 
The system employs various image statistics features (eg.,~mean, standard deviation)
and morphological features (eg.,~number of objects, area). 
The random forest classifier is used.

\begin{figure}[!t]
  \centering
  \includegraphics[width=1.0125\linewidth,height=0.55\linewidth]{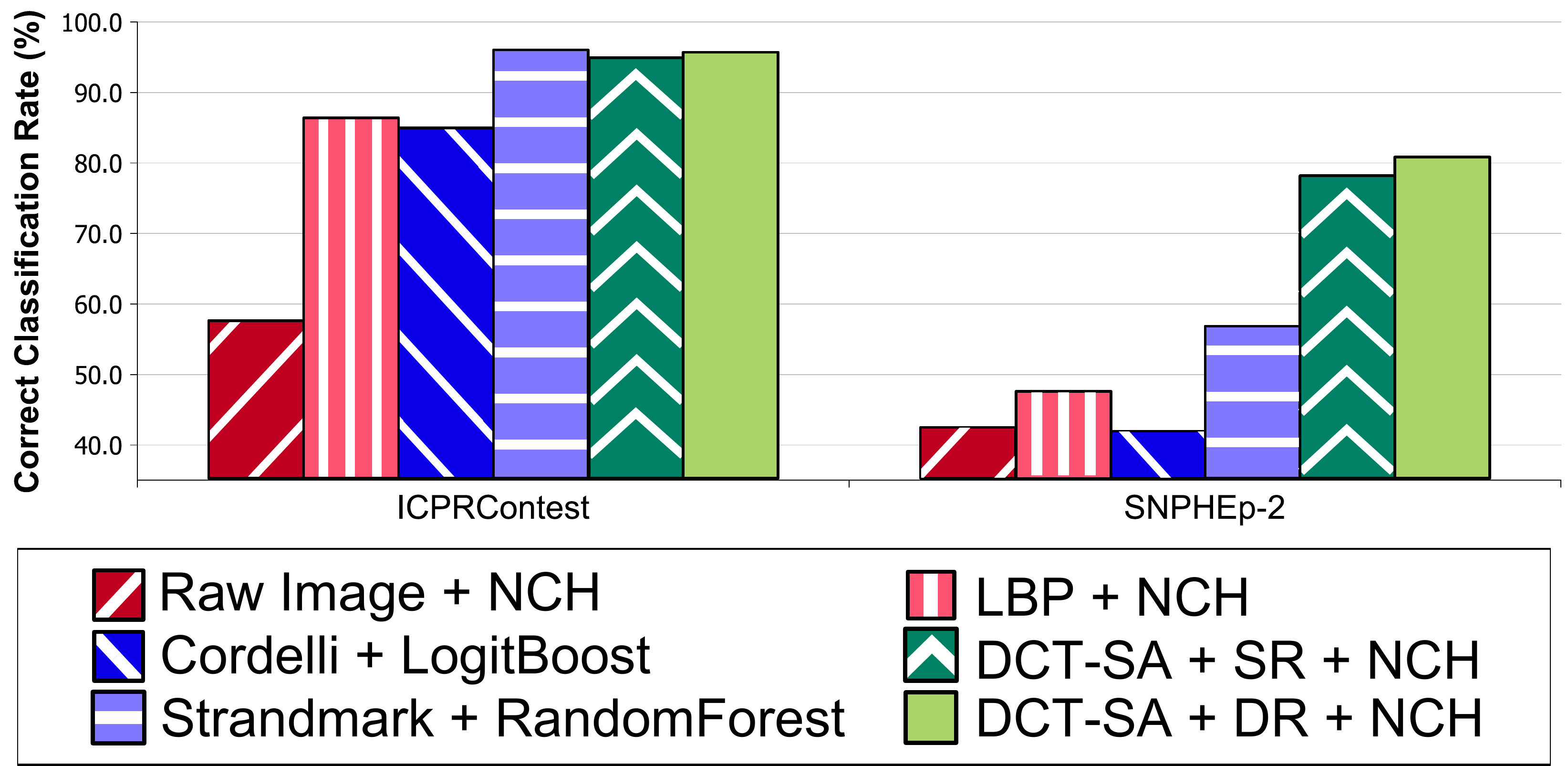} 
  \vspace{-2ex}
  \caption
    {
    Performance comparison of various systems on the ICPRContest and SNPHEp-2 datasets.
    SR =  single region; DR = dual region; NCH = Nearest Convex Hull Classifier.
    }
  \label{fig:system_comparisons}  
\end{figure}

The results are presented in Fig.~\ref{fig:system_comparisons}.
Both the Cordelli and Strandmark systems have reasonable performance on the ICPRContest dataset. 
Strandmark has slightly better performance than the proposed DCT-SA system (96.1\% vs 95.5\%).
However, both the Strandmark and Cordelli systems perform poorly on the more challenging SNPHEp-2 dataset,
while the proposed system has considerably better performance.
This indicates that the descriptors used by Cordelli and Strandmark systems are sensitive to hardware configuration variations.
The poor performance of the Cordelli system on \mbox{SNPHEp-2} can be partly explained from the observation
that LBP also performs poorly on \mbox{SNPHEp-2}. 
As mentioned before, Cordelli uses the LBP channel to compute some of its features.

\balance

\section{Main Findings}
\label{sec:sec_conclusions}

The Indirect Immunofluorescence method on Human Epithelial (\mbox{HEp-2}) cells
is a hallmark method for identifying the presence of Anti-Nuclear Antibodies in clinical pathology tests.
Despite its high sensitivity and the large range of antigens that can be detected,
it has numerous shortcomings, such as being subjective as well as time and labour intensive.
Computer Aided Diagnostic (CAD) systems have been recently developed to address these problems,
which automatically classify a \mbox{HEp-2} cell image into one of the known patterns (eg.,~speckled, homogeneous).
Most of the existing CAD systems use handpicked features to represent a \mbox{HEp-2} cell image,
which may only work in limited scenarios.

In this paper we have proposed a cell classification system comprised of a dual-region codebook-based descriptor
combined with the Nearest Convex Hull Classifier. 
The system splits a cell image into small patches,
which are then grouped into sets representing the inner and edge regions of the cell.
Each region is the described as a histogram of visual words.
To our knowledge, this is the first time codebook-based descriptors are successfully applied 
and thoroughly studied in the domain of cell classification.

We evaluated numerous variants of the descriptor on two publicly available datasets:
ICPR \mbox{HEp-2} cell classification contest dataset and the new \mbox{SNPHEp-2} dataset.
We found that DCT patch-level features in conjunction with soft-assignment/probabilistic encoding of histograms
leads to the highest discrimination performance. 
We also found that imposing the dual-region spatial structure
increases discrimination performance of all codebook-based descriptor variants. 
Furthermore, the experiments show that the proposed system has consistent high performance
and is more robust than two recent CAD systems presented in \cite{Cordelli2011,Strandmark2012}.

We note that the proposed dual-region spatial structure used in this work is intuitive and lacks a theoretical explanation. 
Given the encouraging results, a more complete model of spatial structure could be developed to further increase performance. 

~

\section*{Acknowledgements}

This research was supported by Sullivan Nicolaides Pathology, Australia and NICTA. 
NICTA is funded by the Australian Government as represented by the {\it Department of Broadband, Communications and the Digital Economy}, as well as the Australian Research Council through the {\it ICT Centre of Excellence} program.

~

~

~

\renewcommand{\baselinestretch}{1}\small\normalsize
\small
\bibliographystyle{ieee}
\bibliography{ANA}

\begin{thebibliography}{10}\itemsep=-1pt

\bibitem{Aharon2006}
M.~Aharon, M.~Elad, and A.~Bruckstein.
\newblock {K-SVD:} an algorithm for designing overcomplete dictionaries for
  sparse representation.
\newblock {\em {IEEE} Trans. Signal Processing}, 54(11):4311--4322, 2006.

\bibitem{Bishop2006}
C.~Bishop.
\newblock {\em Pattern Recognition and Machine Learning}.
\newblock Springer, 2006.

\bibitem{Bizzaro1998}
N.~Bizzaro, R.~Tozzoli, E.~Tonutti, A.~Piazza, F.~Manoni, A.~Ghirardello,
  D.~Bassetti, D.~Villalta, M.~Pradella, and P.~Rizzotti.
\newblock Variability between methods to determine {ANA}, {anti-dsDNA} and
  {anti-ENA} autoantibodies: a collaborative study with the biomedical
  industry.
\newblock {\em Journal of Immunological Methods}, 219(1-2):99--107, 1998.

\bibitem{Cevikalp2010}
H.~Cevikalp and B.~Triggs.
\newblock Face recognition based on image sets.
\newblock In {\em IEEE Conf. Computer Vision and Pattern Recognition}, pages
  2567--2573, 2010.

\bibitem{Coates2011}
A.~Coates and A.~Y. Ng.
\newblock The importance of encoding versus training with sparse coding and
  vector quantization.
\newblock In {\em Int. Conf. Machine Learning}, 2011.

\bibitem{Cordelli2011}
E.~Cordelli and P.~Soda.
\newblock Color to grayscale staining pattern representation in {IIF}.
\newblock In {\em International Symposium on {Computer-Based} Medical Systems},
  pages 1--6, 2011.

\bibitem{Elbischger2009}
P.~Elbischger, S.~Geerts, K.~Sander, G.~{Ziervogel-Lukas}, and P.~Sinah.
\newblock Algorithmic framework for {HEp-2} fluorescence pattern classification
  to aid auto-immune diseases diagnosis.
\newblock In {\em {IEEE} International Symposium on Biomedical Imaging: From
  Nano to Macro}, pages 562--565, 2009.

\bibitem{Foggia2010}
P.~Foggia, G.~Percannella, P.~Soda, and M.~Vento.
\newblock Early experiences in mitotic cells recognition on {HEp-2} slides.
\newblock In {\em International Symposium on {Computer-Based} Medical Systems},
  pages 38--43, 2010.

\bibitem{Gurcan2009}
M.~N. Gurcan, L.~E. Boucheron, A.~Can, A.~Madabhushi, N.~M. Rajpoot, and
  B.~Yener.
\newblock Histopathological image analysis: A review.
\newblock {\em {IEEE} Reviews in Biomedical Engineering}, 2:147--171, 2009.

\bibitem{Hiemann2009}
R.~Hiemann, T.~Büttner, T.~Krieger, D.~Roggenbuck, U.~Sack, and K.~Conrad.
\newblock Challenges of automated screening and differentiation of non-organ
  specific autoantibodies on {HEp-2} cells.
\newblock {\em Autoimmunity Reviews}, 9(1):17--22, 2009.

\bibitem{Hsieh2009}
T.~Hsieh, Y.~Huang, C.~Chung, and Y.~Huang.
\newblock {HEp-2} cell classification in indirect immunofluorescence images.
\newblock In {\em Int. Conf. Information, Communications and Signal
  Processing}, pages 1--4, 2009.

\bibitem{Khutlang2010}
R.~Khutlang, S.~Krishnan, R.~Dendere, A.~Whitelaw, K.~Veropoulos, G.~Learmonth,
  and T.~S. Douglas.
\newblock Classification of mycobacterium tuberculosis in images of
  {ZN-stained} sputum smears.
\newblock {\em {IEEE} Trans. Information Technology in Biomedicine},
  14(4):949--957, 2010.

\bibitem{Lazebnik2006}
S.~Lazebnik, C.~Schmid, and J.~Ponce.
\newblock Beyond bags of features: Spatial pyramid matching for recognizing
  natural scene categories.
\newblock In {\em {IEEE} Conference on Computer Vision and Pattern
  Recognition}, volume~2, pages 2169--2178, 2006.

\bibitem{Liu2011}
C.~Liu, J.~Yuen, and A.~Torralba.
\newblock {SIFT} flow: Dense correspondence across scenes and its applications.
\newblock {\em {IEEE} Trans. Pattern Analysis and Machine Intelligence},
  33(5):978--994, 2011.

\bibitem{Lowe2004a}
D.~G. Lowe.
\newblock Distinctive image features from {Scale-Invariant} keypoints.
\newblock {\em International Journal of Computer Vision}, 60:91--110, 2004.

\bibitem{Meroni2010}
P.~L. Meroni and P.~H. Schur.
\newblock {ANA} screening: an old test with new recommendations.
\newblock {\em Annals of the Rheumatic Diseases}, 69(8):1420 --1422, 2010.

\bibitem{Nalbantov2006}
G.~Nalbantov, P.~Groenen, and J.~Bioch.
\newblock Nearest convex hull classification.
\newblock Report EI 2006-50, Erasmus University Rotterdam, Econometric
  Institute, 2006.

\bibitem{Ojala2002}
T.~Ojala, M.~Pietikainen, and T.~Maenpaa.
\newblock Multiresolution gray-scale and rotation invariant texture
  classification with local binary patterns.
\newblock {\em {IEEE} Trans. Pattern Analysis and Machine Intelligence},
  24(7):971--987, 2002.

\bibitem{Perner2002a}
P.~Perner, H.~Perner, and B.~Müller.
\newblock Mining knowledge for {HEp-2} cell image classification.
\newblock {\em Artificial Intelligence in Medicine}, 26:161--173, 2002.

\bibitem{Pham2005}
B.~Pham, S.~Albarede, A.~Guyard, E.~Burg, and P.~Maisonneuve.
\newblock Impact of external quality assessment on antinuclear antibody
  detection performance.
\newblock {\em Lupus}, 14(2):113--119, 2005.

\bibitem{Reddy_TCSVT_2012}
V.~Reddy, C.~Sanderson, and B.~C. Lovell.
\newblock Improved \mbox{foreground} detection via block-based classifier
  cascade with probabilistic decision integration.
\newblock {\em IEEE Transactions on \mbox{Circuits} and Systems for Video
  Technology}, (in press).

\bibitem{Armadillo_2010}
C.~Sanderson.
\newblock Armadillo: An open source {C++} linear \mbox{algebra} library for
  fast prototyping and computationally \mbox{intensive} experiments.
\newblock Technical report, NICTA, 2010.

\bibitem{Sanderson2009}
C.~Sanderson and B.~C. Lovell.
\newblock Multi-region probabilistic histograms for robust and scalable
  identity inference.
\newblock In {\em Lecture Notes in Computer Science (LNCS)}, volume 5558, pages
  199--208, 2009.

\bibitem{Soda2009}
P.~Soda and G.~Iannello.
\newblock Aggregation of classifiers for staining pattern recognition in
  antinuclear autoantibodies analysis.
\newblock {\em {IEEE} Trans. Information Technology in Biomedicine},
  13(3):322--329, 2009.

\bibitem{Strandmark2012}
P.~Strandmark, J.~Ul\'{e}n, and F.~Kahl.
\newblock Hep-2 staining pattern classification.
\newblock In {\em Int. Conf. Pattern Recognition}, 2012.

\bibitem{Tommasi2008}
T.~Tommasi, F.~Orabona, and B.~Caputo.
\newblock Discriminative cue integration for medical image annotation.
\newblock {\em Pattern Recognition Letters}, 29(15):1996--2002, 2008.

\bibitem{Tropp2010}
J.~Tropp and S.~Wright.
\newblock Computational methods for sparse solution of linear inverse problems.
\newblock {\em Proceedings of the {IEEE}}, 98(6):948 --958, 2010.

\bibitem{Gemert2010}
J.~van Gemert, C.~Veenman, A.~Smeulders, and J.~Geusebroek.
\newblock Visual word ambiguity.
\newblock {\em {IEEE} Trans. Pattern Analysis and Machine Intelligence},
  32(7):1271--1283, 2010.

\bibitem{Wiik2010}
A.~S. Wiik, M.~{Høier-Madsen}, J.~Forslid, P.~Charles, and J.~Meyrowitsch.
\newblock Antinuclear antibodies: A contemporary nomenclature using {HEp-2}
  cells.
\newblock {\em Journal of Autoimmunity}, 35:276--290, 2010.

\bibitem{Wong_IJCNN_2012}
Y.~Wong, M.~T. Harandi, C.~Sanderson, and B.~C. Lovell.
\newblock On robust biometric identity verification via sparse encoding of
  faces: Holistic vs local approaches.
\newblock In {\em IEEE International Joint Conference on Neural Networks},
  pages 1762--1769, 2012.

\bibitem{Yang2009}
J.~Yang, K.~Yu, Y.~Gong, and T.~Huang.
\newblock Linear spatial pyramid matching using sparse coding for image
  classification.
\newblock In {\em IEEE Conf. Computer Vision and Pattern Recognition}, pages
  1794--1801, 2009.

\end{thebibliography}

\end{document}